\documentclass[runningheads]{llncs}
\usepackage[T1]{fontenc}
\usepackage{graphicx}
\usepackage{cite}
\usepackage{booktabs} 
\usepackage{multirow}
\usepackage{amssymb}
\usepackage{array}
\usepackage{hyperref}
\usepackage{marvosym}

\usepackage{color}

\begin{document}
\title{Common Sense Enhanced Knowledge-based Recommendation with Large Language Model}
%
\author{Shenghao Yang\inst{1} \and
Weizhi Ma\inst{2}\textsuperscript{ \Letter} \and Peijie Sun\inst{1} \and Min Zhang\inst{1}\textsuperscript{ \Letter} \and
Qingyao Ai\inst{1} \and Yiqun Liu\inst{1} \and Mingchen Cai\inst{3}}
\authorrunning{S. Yang et al.}

\institute{DCST, Tsinghua University, Beijing 100084, China \\
\email{ysh21@mails.tsinghua.edu.cn, sun.hfut@gmail.com, \{z-m,aiqy,yiqunliu\}@tsinghua.edu.cn} \\ \and
AIR, Tsinghua University, Beijing 100084, China \\
\email{mawz@tsinghua.edu.cn} \\ \and
Meituan Inc, Beijing, China 
\email{caimingchen@meituan.com}}
\maketitle              
\begin{abstract}
Knowledge-based recommendation models effectively alleviate the data sparsity issue leveraging the side information in the knowledge graph, and have achieved considerable performance. Nevertheless, the knowledge graphs used in previous work, namely metadata-based knowledge graphs, are usually constructed based on the attributes of items and co-occurring relations (e.g., also buy), in which the former provides limited information and the latter relies on sufficient interaction data and still suffers from cold start issue. Common sense, as a form of knowledge with generality and universality, can be used as a supplement to the metadata-based knowledge graph and provides a new perspective for modeling users' preferences. Recently, benefiting from the emergent world knowledge of the large language model, efficient acquisition of common sense has become possible. In this paper, we propose a novel knowledge-based recommendation framework incorporating common sense, CSRec, which can be flexibly coupled to existing knowledge-based methods. Considering the challenge of the knowledge gap between the common sense-based knowledge graph and metadata-based knowledge graph, we propose a knowledge fusion approach based on mutual information maximization theory. Experimental results on public datasets demonstrate that our approach significantly improves the performance of existing knowledge-based recommendation models.

\end{abstract}
\section{Introduction}
Recommender systems have been widely applied to address the issue of information overload in various internet services, exhibiting promising performance in scenarios such as e-commerce platforms~\cite{hou2023deep} and video recommendations~\cite{covington2016deep}. 
traditional recommender system techniques are typically based on collaborative filtering algorithms~\cite{sarwar2001item}, which heavily rely on the availability of substantial user-item interaction data. Nevertheless, real-world scenarios usually present a sparse interaction matrix between users and items, accompanied by the emergence of new users and items, leading to issues of data sparsity and the cold start issue~\cite{sun2019research}.


To address the challenges above, knowledge-based approaches have been proposed to alleviate the reliance on substantial interaction data by incorporating rich side information from knowledge graphs into the user and item representation modeling process~\cite{guo2020survey}. 
While existing knowledge-based recommenders have exhibited considerable effectiveness, there are certain limitations inherent in the knowledge graphs used in previous methods. Typically, these knowledge graphs are constructed based on metadata from the dataset, referred to as metadata-based knowledge graphs. The utilized metadata may include item attributes
However, the information derived from attribute-based item relations is inherently limited in scope. Another form of metadata involves co-occurrence relations extracted from user interaction data, such as ``also bought” and ``also view” relations~\cite{ai2018learning}. Although these interaction-derived item relations offer valuable insights, they remain highly reliant on extensive interaction data, thereby still susceptible to challenges arising from data sparsity and the cold start issue. 
Common sense, widely acknowledged and universally accepted, presents two advantages when utilized in constructing knowledge graphs. Firstly, as a form of knowledge with generality and ubiquity, common sense effectively supplements the information present in metadata. Secondly, common sense is data-agnostic, eliminating the need for reliance on substantial interaction data for extraction. Thirdly, common sense can capture intuitive human insights, providing a new perspective for modeling user preferences.
Nevertheless, integrating common sense into the construction of recommendation knowledge graphs has several challenges. On one hand, common sense is inherently unstructured, presenting difficulties in organization and integration. On the other hand, synthesizing common sense requires the design of intricate annotation schemes, incurring significant manual labor costs. 

Recently, large language models~\cite{openai2023gpt4,touvron2023llama} have a revolutionary advancement in various natural language processing tasks, showcasing remarkable performance improvements and demonstrating a rich common sense during task-solving processes. Based on this observation, we propose to harness common sense from large language models to construct the knowledge graph for recommendation.
Specifically, the common sense of the large language model is leveraged to analyze complementary and substitute relations of item categories, such as "I bought a phone, please recommend some complementary item categories."

In this paper, we propose a knowledge-based recommender system framework, CSRec, which incorporates \textbf{C}ommon \textbf{S}ense knowledge into the construction of knowledge graphs tailored for \textbf{Rec}ommender systems. 
Specifically, we integrate the relations between categories based on the common sense insights from the large language model and construct a knowledge graph. 
Considering the distinct sources and dimensions of the two knowledge graphs, direct merging leads to challenges for graph embedding learning. To address the above issue, we propose a knowledge fusion method based on mutual information maximization (MIM)~\cite{logeswaran2018efficient} theory. 

In the experiment, We evaluate the performance of CSRec on public datasets to assess the efficacy of our approach. The experiments demonstrate that integrating a common sense-based graph into a knowledge-based recommender, coupled with the MIM-based knowledge fusion strategy, significantly improves the performance of the vanilla methods. Moreover, in cold start scenarios, CSRec achieved a notable improvement in the performance of cold users and cold items.
The contributions of our work can be summarized as follows:

\begin{itemize}
    \item We present the effectiveness of incorporating common sense derived from the large language model (LLM) into knowledge-based recommender systems.
    \item We propose a novel framework, CSRec, which constructs an LLM-based common sense knowledge graph and integrates it into the recommendation framework with a mutual information maximization (MIM)-based knowledge fusion approach.
    \item Experimental results demonstrate that our approach effectively improves the recommendation performance and it is flexible to be coupled with various knowledge-based models.
\end{itemize}

\section{Related work}
\subsection{Knowledge-based Recommendation}
Knowledge-based recommender systems leverage side information from knowledge graphs to effectively address the issue of data sparsity. Based on the utilization of knowledge graph information, existing works can be categorized into three categories: embedding-based methods, path-based methods, and unified methods

\textbf{Embedding-based methods} adopt graph embedding techniques to learn representations of entities and relationships in the knowledge graph. The entity representation obtained is then used to enhance the representation of corresponding items or users. CKE~\cite{zhang2016collaborative} constructs a knowledge graph containing items and attributes and adopts the TransR algorithm to learn representations of items and attributes in the knowledge graph. The representation of an item in the knowledge graph combined with its origin embedding is used for obtaining user-item preference scores.
CFKG~\cite{ai2018learning} incorporates users as entities into the knowledge graph. It defines a scoring function to calculate the score between two entities given a relationship. This function is utilized for both training graph embeddings and computing scores for a user-item pair given the "buy" relationship in evaluation.

\textbf{Path-based methods} leverage connectivity patterns in the user-item graph for recommendation, focusing on exploring the connected similarity between users or items in the graph. Hete-MF~\cite{yu2013collaborative} calculates similarity scores among items through multiple meta-paths and uses these scores to refine item representations.

\textbf{Unified methods} inspired by embedding propagation, integrate both semantic and connectivity information from the knowledge graph. KGCN~\cite{wang2019knowledge} aggregates information from neighboring nodes in the knowledge graph using Graph Convolutional Networks (GCN) to derive the final representation of items. KGAT~\cite{wang2019kgat} further incorporates users into the knowledge graph and creates a collaborative knowledge graph (CKG). It then adopts a Graph Neural Network (GNN) model for embedding propagation on user and item representations.

The above approaches rely on metadata-based knowledge graphs and our proposed method proposes a knowledge graph based on common sense, providing a supplement to metadata-based knowledge graphs. Therefore, our approach can be considered orthogonal to existing methods and can be seamlessly integrated into any of the aforementioned models.

\subsection{Large Language Model for  Recommendation}
The emergence of large language models (LLMs)~\cite{brown2020language,openai2023gpt4,touvron2023llama,zhang2022opt} has revolutionized the field of natural language processing (NLP), showcasing remarkable reasoning capabilities and extensive world knowledge. The application of LLMs to recommender tasks has been widely explored, with research efforts primarily falling into two main lines of inquiry.

One approach aims to harness the reasoning capabilities and knowledge of LLMs into various stages of the recommendation process. Liu et al.~\cite{liu2023chatgpt} utilized ChatGPT for five recommendation tasks in zero-shot and few-shot settings, achieving impressive results in tasks such as explanation generation. However, the model without training on user interaction data demonstrated suboptimal performance in topN and sequential recommendation scenarios. P5~\cite{zhang2023recommendation} adopted transforms recommendation data and tasks into natural language formats as inputs for T5. Through fine-tuning with generative task objectives, the model exhibited enhanced performance across five recommendation tasks.

The second approach explores the combination of recommendation models with LLMs. ChatRec~\cite{gao2023chatrec} proposes to use recommendation models for extensive item recall, followed by leveraging LLMs for reranking the candidate items. TALLRec~\cite{Bao_2023} efficiently utilizes parameter-efficient methods, namely LoRA, for fine-tuning LLaMA, enabling the model to learn recommendation-related knowledge and apply it to the item reranking. InstructGPT~\cite{zhang2023recommendation} derives instruction samples from user interaction data and reviews data for instructing fine-tuning FLANT5, resulting in an LLM tailored for recommendation tasks. ONCE~\cite{liu2023once} adopts ChatGPT as a data augmenter to enhance the representation of users and news, significantly improving the performance of the news recommendation.

However, existing methods exhibit significant limitations in both performance and effectiveness. In this paper, we propose leveraging the common sense of LLMs for constructing a knowledge graph used in recommendation. Considering the process is performed offline, it maximizes the utilization of LLM's knowledge while mitigating efficiency concerns.

\section{Method}
In this section, we begin by formalizing the knowledge-based recommender system, utilizing the primary symbols outlined in Table 1. Subsequently, we present the proposed common sense-based recommendation model framework, CSRec, encompassing the construction of a common sense graph and the recommendation process based on common sense.

\begin{figure}[t]
\includegraphics[width=\textwidth]{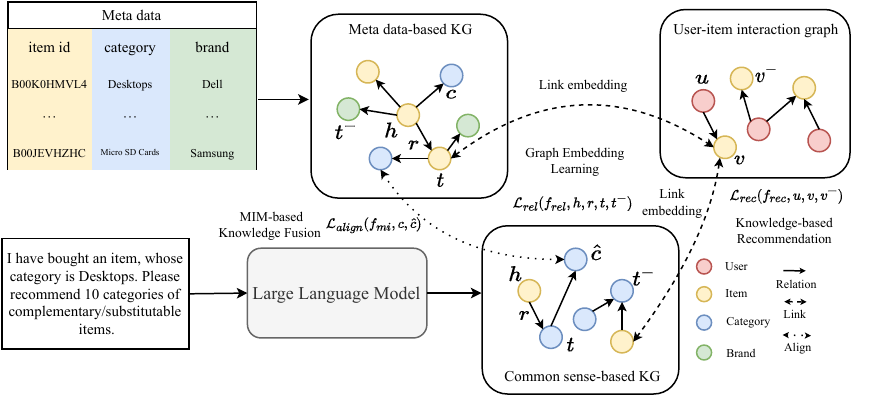}
\caption{The overall framework of Common sense-based Recommendation.} \label{fig1}
\end{figure}

\subsection{Problem Definition}
\begin{table}[t]
\begin{center}
\caption{Notations.}
  \label{tab: notations}
\begin{tabular}{ll}
\toprule
Notations & Descriptions \cr
\midrule
$u_i$ & User $i$ \cr
$v_j$ & Item $j$ \cr
$e_k$ & Entity $k$ in the knowledge graph \cr
$r_e$ & Relation between two entities $(e_i, e_j)$ in the knowledge graph \cr
${U}=\{u_1,u_2,...,u_m\}$ & User set \cr
${V}=\{v_1,v_2,...,v_n\}$ & Item set \cr
$\mathcal{V}=\{e_1,e_2,...,e_p\}$ & Item set \cr
$\mathcal{E}=\{r_1,r_2,...,r_q\}$ & Item set \cr
$E_u \in \mathbb{R}^{|{U}| \times d}$ & User embedding \cr
$E_v \in \mathbb{R}^{|{V}| \times d}$ & Item embedding \cr
$E_e \in \mathbb{R}^{|\mathcal{V}| \times d}$ & Entity embedding \cr
$E_r \in \mathbb{R}^{|\mathcal{E}| \times d}$ & Relation embedding \cr
$f_{rec}$ & Score function for user-item pair $(u_i,v_j)$ \cr
$f_{rel}$ & Score function for triplet $(e_i,r_k,e_j)$ \cr
$f_{MI}$ & Score function for the mutual information of entity pair $(e,\hat{e})$ \cr

\bottomrule
\end{tabular}
\end{center}
\end{table}
In this section, we begin by formalizing the entire knowledge-based recommendation problem. The sets of users and items are respectively denoted as $U = \{u_1, u_2, ..., u_n\}$ and $V = \{v_1, v_2, ..., v_m\}$. The knowledge graph is a heterogeneous graph, denoted by $\mathcal{G} = (\mathcal{V}, \mathcal{E})$. $R$ denotes the set of all relation triples in the knowledge graph. Each triple is represented as $(h, r, t)$, where $h$, $r$, and $t$ denote the head node, relation edge, and tail node, respectively.

In the knowledge graph utilized for recommendation, nodes typically encompass users, items, and item attributes, collectively referred to as entities $e$. The objective of the recommendation task is to learn a scoring function $f_{rec}(u, v)$ that predicts the preference score $y_{u,v}$ of user $u$ for item $v$, ultimately facilitating the ranking of items based on these scores. To achieve this, we obtain representations $u$ and $v$ for users and items, respectively. Additionally, leveraging the knowledge graph allows us to obtain representations $e$ and $r$ for all entities and relations. Through certain mapping methods, users and items are associated with corresponding entities on the graph, and all representation information is utilized for the scoring function. The notations are shown in Table~\ref{tab: notations}.
\subsection{Framework Overview}
In this section, we present our prompt common sense-based recommendation framework CSRec, the overall framework is shown in Fig~\ref{fig1}. Our principal concept involves integrating two common sense-related modules into existing knowledge-based recommender systems: the common sense graph and the knowledge graph fusion based on mutual information maximization. It is noteworthy that our approach is model-agnostic, enabling seamless integration with any knowledge-based recommendation model.

Current knowledge-based recommendation models primarily focus on the joint optimization of user-item representations and entity relation representations within the knowledge graph. They achieve this by linking items and entities through a mapping function, allowing the optimization objective to maximize recommendation performance. Our proposed framework retains the same optimization objective, simultaneously learning representations for both knowledge graphs and merging the two. The framework comprises two main components: Common Sense-based KG construction and Recommendation with Common Sense-based KG.
\subsection{Common Sense-based KG construction}
In this section, we present the methodology for constructing a common sense-based Knowledge Graph (KG). We begin by providing definitions for entities and relations within the KG. Subsequently, we illustrate the approach for analyzing relations between entities. Finally, we elucidate how to integrate the analysis results onto the edges of the KG.

\subsubsection{Entity and Relation}
Considering that common sense is a form of unstructured knowledge, and knowledge graphs represent a structured data format, the construction process can be viewed as the transformation of common sense into structured data. Initially, we need to define entities and relations within the knowledge graph. Since common sense is highly abstract and suited for handling broad concepts, we focus on two intuitive concepts relevant to recommendation tasks: complementarity and substitutes. Complementarity refers to the scenario where two items can be used together, such as a phone and a phone case. At the same time, substitutes imply that two items can replace each other and serve similar functions, for example, a tablet and an e-reader. These relations align well with human intuition and are conducive to common sense judgment.

The most straightforward approach for entities is to use items as entities. However, in reality, item information is precise and diverse, deviating from the high-level abstraction inherent in common sense. Simultaneously, the sheer volume of items introduces a high time complexity for analyzing relations between them. Therefore, we consider using item categories as entities. Categories inherently represent a higher-level clustering of items, embodying more generalized information about items. Consequently, we can initially establish relations between categories, naturally extending to relations between items and categories. Given that the quantity of categories is significantly smaller than that of items, this approach enables a more efficient graph construction process.
\subsubsection{Relation Analysis}
To obtain complementary and substitutable categories corresponding to a specific item category, a straightforward approach involves manual annotation. However, this process incurs significant human labor costs in practice. Recently, with the observed knowledge proficiency of large language models approaching human levels, we propose utilizing a large language model for the analysis of category relations. For a dataset with a collection of categories $C = \{c_1, c_2, ..., c_{|C|}\}$, we construct two types of prompts as inputs for the large language model, as illustrated below:

\begin{quote}
"I have bought an item, whose category is \{item category\}. Please recommend 10 categories of complementary/substitutable items. Output format: One category one line, without any explanations."
\end{quote}

Based on the feedback from the large language model, we can obtain 10 pseudo-item categories for each item category. Here, "pseudo" denotes that the categories returned by the large language model may not necessarily belong to the set $C$.

\subsubsection{Analysis result integration}
In the preceding section, we obtained 10 pseudo-categories with complementary and substitutable relations for each category. However, to maintain consistency with the categories in the dataset, we need to map these pseudo-categories back to the dataset's category set. Specifically, we employ a text retrieval model, BM25, to accomplish category mapping. The dataset's category set serves as the corpus, with each category acting as a document, and all pseudo-categories as queries. By retrieving the most relevant true category for each pseudo-category, we establish the mapping back to the dataset's category set. This process allows us to construct edges between category entities representing complementary and substitutable relations.

Furthermore, we extend the graph's richness by establishing edges between items corresponding to categories and categories involved in such relations. This augmentation serves to enhance the overall information content within the graph.
\subsection{Recommendation with Common Sense-based KG}
\subsubsection{Knowledge-based User and Item Repersentation}
In this stage, given a user and an item, their representations $\mathbf{u}$ and $\mathbf{v}$ are obtained through the following formulas:
\begin{equation}
\mathbf{u} = \lambda_{u_1} E_u(u) + \lambda_{u_2} E_e(f_{u,e}(u)) + \lambda_{u_3} \hat E_e(f_{u,e}(u)),
\end{equation}
\begin{equation}
\mathbf{v} = \lambda_{v_1}E_u(v) + \lambda_{v_2} E_e(f_{v,e}(v)) + \lambda_{v_3} \hat E_e(f_{v,e}(v)),
\end{equation}
Where $E_u, E_v, E_e, \hat E_e$ represent the embeddings of entities in the user, item, metadata-based knowledge graph, and common sense-based knowledge graph, respectively. $f_{u,e}$ and $f_{v,e}$ denote the mapping functions from users and items to entities. The values of $\lambda_{u_i}$ and $\lambda_{v_i}$ are selected from $\{0,1\}$, controlling whether to link users or items to entities to obtain their representations in the knowledge graph.

User preferences for items are calculated through a scoring function, and without loss of generality, we use the dot product as the scoring function:
\begin{equation}
y_{u,v} = f_{rec}(u,v) = \mathbf{u}^T \mathbf{v}.
\end{equation}

During the training phase, we optimize the scoring function using Bayesian Personalized Ranking (BPR) loss:

\begin{equation}
\mathcal{L}_{rec} = -\sum_{u\in U} \log \sigma(y_{u,v} - y_{u,v^-}).
\end{equation}

In the evaluation phase, we directly rank all candidate items based on $y_{u,v}$.

\subsubsection{Knowledge Graph Embedding Learning}
In this section, based on the knowledge graph, entity and relation representations are optimized through a pair-wise loss, i.e.,
\begin{equation}
\mathcal{L}_{rel} = -\sum_{(h,r,t)\in R} \log \sigma(f_{rel}(h,r,t) - f_{rel}(h,r,t^-)),
\end{equation}
Where $f_{rel}$ is a scoring function used to calculate the score for a triple, and any graph embedding method can be employed. Taking transE as an example:
\begin{equation}
f_{rel}(h,r,t) = -||\mathbf{h} + \mathbf{r} - \mathbf{t}||^2.
\end{equation}

It is essential to note that in our framework, the two knowledge graphs, metadata-based and common sense-based, optimize their respective graph embeddings independently. 

\subsubsection{Mutual Information Maximization based Knowledge fusion}
In our proposed CSRec framework, two knowledge graphs from different knowledge sources are included: the metadata-based knowledge graph and the common sense-based knowledge graph. To merge these two knowledge graphs, the most straightforward approach would be to directly combine their nodes and edges. However, we argue that such a hard merging method may not be conducive to the learning process of graph embeddings, given the substantial gap in knowledge spaces between the two graphs. One originates from attribute information and item co-occurrence statistics within a dataset, while the other stems from highly abstract common sense.

We propose a relatively soft merging approach, considering that the relations obtained through common sense are between item categories. As categories represent overlapping entity sets in both knowledge graphs, we use categories as a bridge to facilitate the fusion of the two knowledge graphs.


Specifically, considering the same category entity in two knowledge graphs, denoted as $c$ and $\hat{c}$, they can be regarded as two views of the same category in different knowledge spaces. Our optimization objective is to maximize the correlation between them, typically measured by the mutual information between the two, defined as:
\begin{equation}
I(c, \hat{c}) = H(c) - H(c | \hat{c}) = H(\hat{c}) - H(\hat{c} | c).
\end{equation}

Therefore, our optimization objective becomes learning a function $f_{mi}$ with $c$ and $\hat{c}$ as inputs to maximize $I(c, \hat{c})$. 
When $f_{mi}$ involves a neural network structure, maximizing mutual information is often challenging. Previous works usually optimize a lower bound of $I(c, \hat{c})$, with InfoNCE being the most widely used lower bound, defined as:
\begin{equation}
I(c, \hat{c}) \ge \mathbb{E}_{p(c, \hat{c})}[f_{mi}(c, \hat{c}) - \mathbb{E}_{q(\widetilde{\mathcal{C}})}[\log \sum_{\widetilde{c} \in \widetilde{\mathcal{C}}} \exp f_{mi}(c, \widetilde{c})]] + \log|\widetilde{\mathcal{C}}|,
\end{equation}
Where $\widetilde{\mathcal{C}}$ is the set of all categories. The optimization process based on the above equation is commonly referred to as contrastive learning, and it can be implemented using cross-entropy loss:
\begin{equation}
\mathcal{L}_{align} = -\sum_{c \in \widetilde{\mathcal{C}}} \log\frac{\exp(f_{mi}(c, \hat{c})/\tau)}{\sum_{\widetilde{c} \in \widetilde{\mathcal{C}}} \exp(f_{mi}(c, \widetilde{c})/\tau)},
\end{equation}
Where $\tau$ is the temperature coefficient for the cross-entropy loss. Without loss of generality, we use the dot product of the vector representations of $c$ and $\hat{c}$ as the scoring function $f_{mi}$.

Finally, we jointly optimize the objectives of the three modules:
\begin{equation}
\mathcal{L} = \mathcal{L}_{rec} + \mathcal{L}_{rel} + \hat{\mathcal{L}}_{rel} + \mathcal{L}_{align}.
\end{equation}

Note that there are two graph embedding learning losses in the formulation, as we optimize the graph embeddings separately for the metadata-based knowledge graph and the common sense-based knowledge graph.

\section{Experiment}
In this section, we first introduce the experiment setting and then present the experimental results and analysis. 
\subsection{Experiment Setting}
\subsubsection{Datasets}
We evaluated the performance of our proposed CSRec framework on two public datasets, namely, Amazon Electronics and Amazon Office Products. Following previous practices, we filtered out users and items with fewer than 5 interactions. The statistical information for the preprocessed versions of the two datasets is presented in table~\ref{tab:data_tabel} and table~\ref{tab:kg_tabel}.
\begin{table}[t]
\tabcolsep=6pt
\renewcommand\arraystretch{1.2}
\begin{center}
\caption{Statistical information of datasets}
  \label{tab:data_tabel}
\begin{tabular}{lllll}
\toprule
Dataset & Users & Items & Inters. & Avg. n \cr
\midrule
Electronics & 177,578 & 59,236 & 1,551,783 & 8.73 \cr
Office & 4,906 & 2,421 & 53,258 & 10.86 \cr
\bottomrule
\end{tabular}
\end{center}
\end{table}

\begin{table}[t]
\tabcolsep=6pt
\renewcommand\arraystretch{1.2}
\begin{center}
\caption{Statistical information of Knowledge graph}
  \label{tab:kg_tabel}
\begin{tabular}{llllll}
\toprule
Dataset & Version & Entities & Relations & Triples \cr
\midrule
\multirow{3}{*}{\makebox[0.13\textwidth][c]{Electronics}} 
& Metadata based & 303,994 & 6 & 2,865,744  \cr
& Common sense based & 62,372 & 6 & 553,877  \cr
& Merge & 305,237 & 10 & 3,419,621  \cr
\midrule
\multirow{3}{*}{\makebox[0.13\textwidth][c]{Office}} & Metadata based & 50,687 & 6 & 239,611  \cr
& Common sense based & 2700 & 6 & 35,243  \cr
& Merge & 50,687 & 10 & 274,854  \cr
\bottomrule
\end{tabular}
\end{center}
\end{table}
\subsubsection{Baselines}
As our proposed CSRec can be flexibly applied to any knowledge-based recommendation model, we conducted comparisons between the original models and the models enhanced with CSRec. Additionally, we compared them with several CF-based recommendation models.

The CF-based recommendation models include:
\begin{itemize}
    \item \textbf{BPR}~\cite{rendle2012bpr} is a matrix factorization model based on the Bayesian Personalized Ranking objective function.
    \item \textbf{NeuMF}~\cite{he2017neural} is a neural network-based matrix factorization model that models user-item relationships.
\end{itemize}

The Knowledge-based models include:
\begin{itemize}
    \item \textbf{CKE}~\cite{zhang2016collaborative} constructs a knowledge graph based on item attributes and jointly optimizes the recommendation loss and graph embedding learning loss.
    \item \textbf{CFKG}~\cite{ai2018learning} merges the user-item interaction graph and item attribute graph, optimizing a triplet scoring function to learn representations of entities and relationships. It utilizes the scores under the "buy" relationship for ranking candidate items.
\end{itemize}

\subsubsection{Metrics}
We use two widely used evaluation metrics, HR@K and nDCG@K, to evaluate the performance of all models in the TopN recommendation task. K is set to 10 and 50. Following previous works, we split the dataset into train, valid, and test sets with a ratio of \{0.8:0.1:0.1\}.
\subsubsection{Implementation Details}
We implement the CSRec framework and baseline models using the RecBole library. As for the large language model, we use ChatGPT. For CSRec, the temperature of align loss is searched in the range of \{0.5, 0.3, 0.1, 0.05, 0.03, 0.01\}. For all models, we use Adam optimizer and carefully search for hyperparameters, with a batch size of 2048 and early stopping with the patience of 10, using nDCG@10 as the indicator. We tune the learning rate in \{0.0003, 0.001, 0.003, 0.01\} and the embedding dimension in \{64, 128, 300\}. The code is available at \url{https://github.com/ysh-1998/CSRec}.
\subsection{Overall Experimental Results}
\begin{table}[t]
\begin{center}
\tabcolsep=6pt
\renewcommand\arraystretch{1.2}
\caption{Performance comparison between CSRec and base models.}
  \label{tab:tab3}
\begin{tabular}{lllllll}
\toprule
\makebox[0.13\textwidth][c]{Dataset} & Model     & HR@10 & HR@50 & nDCG@10 & nDCG@50 \cr
\midrule
\multirow{10}{*}{\makebox[0.13\textwidth][c]{Electronics}} 
& BPR&0.2273&0.4425&0.1294&0.1759 \cr
& NeuMF&0.2096&0.4238&0.1179&0.1639 \cr
\cmidrule(lr){2-2} \cmidrule(lr){3-6} 
& CKE$_{meta}$       &0.2288&0.4480&0.1303&0.1777 \cr
& CKE$_{cs}$       &0.2266&0.4450&0.1289&0.1761 \cr
& CSRec(CKE)  &\textbf{0.2343$^{**}$}&\textbf{0.4546$^{**}$}&\textbf{0.1331$^{*}$}&\textbf{0.1807$^{*}$} \cr
\cmidrule(lr){2-2} \cmidrule(lr){3-6} 
& CFKG$_{meta}$      &0.2712&0.4949&0.1568&0.2054 \cr
& CFKG$_{cs}$      &0.2287&0.4437&0.1293&0.1758 \cr
& CSRec(CFKG) 
&\textbf{0.2766$^{**}$}&\textbf{0.5025$^{**}$}&\textbf{0.1615$^{**}$}&\textbf{0.2106$^{**}$} \cr
\bottomrule
\end{tabular}
\begin{tabular}{lllllll}
\toprule
\makebox[0.13\textwidth][c]{Dataset} & Model     & HR@10 & HR@50 & nDCG@10 & nDCG@50 \cr
\midrule
\multirow{10}{*}{\makebox[0.13\textwidth][c]{Office}} 
& BPR&0.1233&0.3252&0.0635&0.1050 \cr
& NeuMF&0.1360&0.3513&0.0634&0.1082 \cr
\cmidrule(lr){2-2} \cmidrule(lr){3-6} 
& CKE$_{meta}$       &0.1392&0.3368&0.0676&0.1088 \cr
& CKE$_{cs}$       &0.1339&0.3366&0.0662&0.1083 \cr
& CSRec(CKE)  &\textbf{0.1423$^{*}$}&\textbf{0.3470$^{**}$}&\textbf{0.0698$^{*}$}&\textbf{0.1121$^{**}$} \cr
\cmidrule(lr){2-2} \cmidrule(lr){3-6} 
& CFKG$_{meta}$      &0.1394&0.3437&0.0684&0.1105 \cr
& CFKG$_{cs}$      &0.1307&0.3482&0.0594&0.1047 \cr
& CSRec(CFKG) 
&\textbf{0.1488$^{**}$}&\textbf{0.3592$^{**}$}&\textbf{0.0742$^{**}$}&\textbf{0.1176$^{**}$} \cr
\bottomrule
\end{tabular}
\end{center}
\end{table}
The overall experimental results are presented in Table 1, and several observations can be derived. Firstly, leveraging the rich side information provided by the knowledge graph, knowledge-based recommendation models consistently outperform models based on collaborative filtering (CF). This demonstrates the effectiveness of the knowledge graph's information in alleviating data sparsity issues. Secondly, by incorporating users into the knowledge graph, the CFKG-based model achieves overall better performance compared to the CKE-based model. Thirdly, the models using a metadata-based knowledge graph tend to yield better results than models using a common sense-based knowledge graph. This is because the former includes rich item relationships based on user interaction information, while the latter only contains common sense information on item categories. It's more suitable for supplementing the knowledge lacking in the former rather than being used independently.

Our proposed CSRec framework achieves the most significant performance improvement on both base models and outperforms CF-based methods significantly. That indicates the effectiveness of incorporating common sense to complement metadata-based knowledge graphs and demonstrates the effectiveness of our proposed knowledge fusion method based on mutual information maximization for integrating common sense into existing models.

\subsection{Ablation Study}
\begin{figure}[t]
\includegraphics[width=\textwidth]{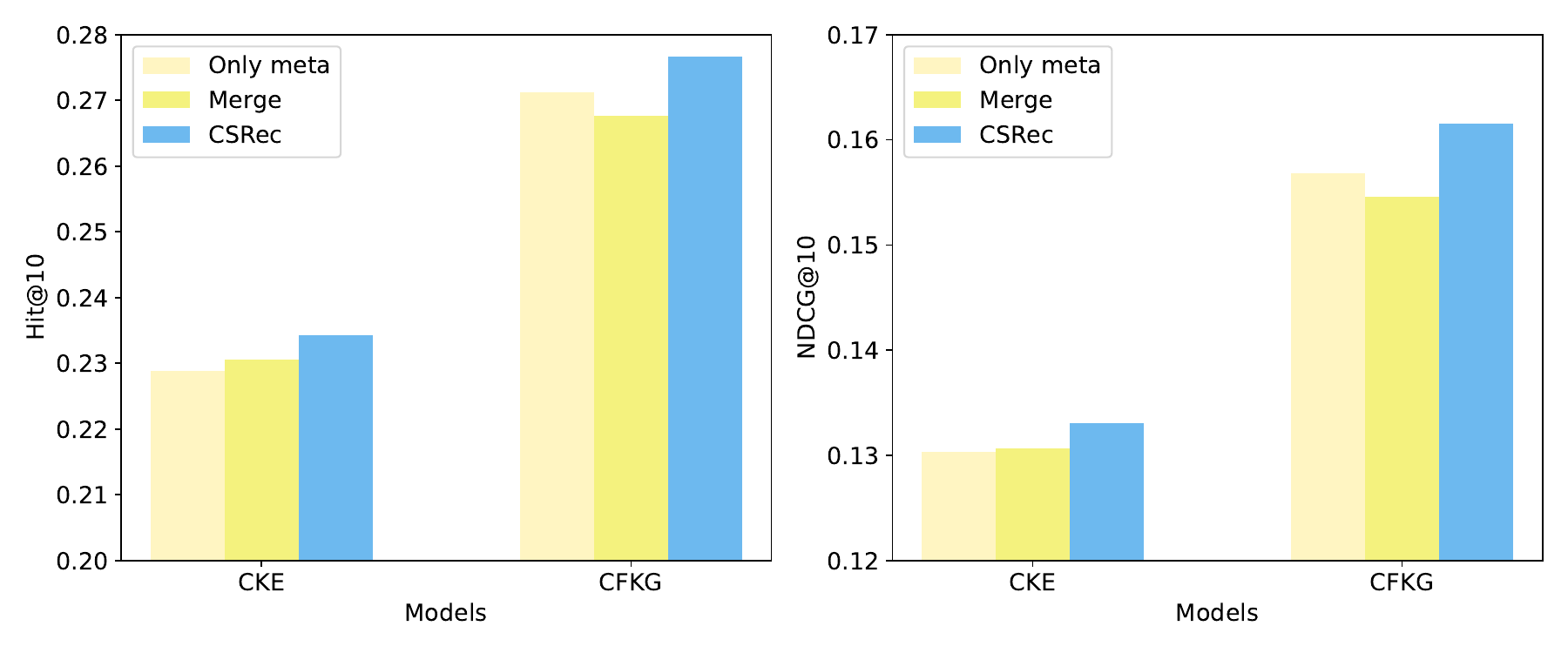}
\caption{Ablation study results.} \label{fig2}
\end{figure}
When simultaneously applying two knowledge graphs to the recommendation model, an intuitive idea is to directly merge the two graphs. This is a variant of our proposed CSRec method without the knowledge fusion component based on mutual information maximization. Figure~\ref{fig2} compares the experimental results of this variant with our proposed method based on different base models. From the results, it can be observed that merge mothod consistently performs worse than our CSRec. We speculate that this is because the knowledge sources and information dimensions of the two knowledge graphs exhibit significant gaps, and the simplistic merging approach led to negative effects on graph embedding learning.
\subsection{Cold Start Performance}
\begin{figure}[t]
\includegraphics[width=\textwidth]{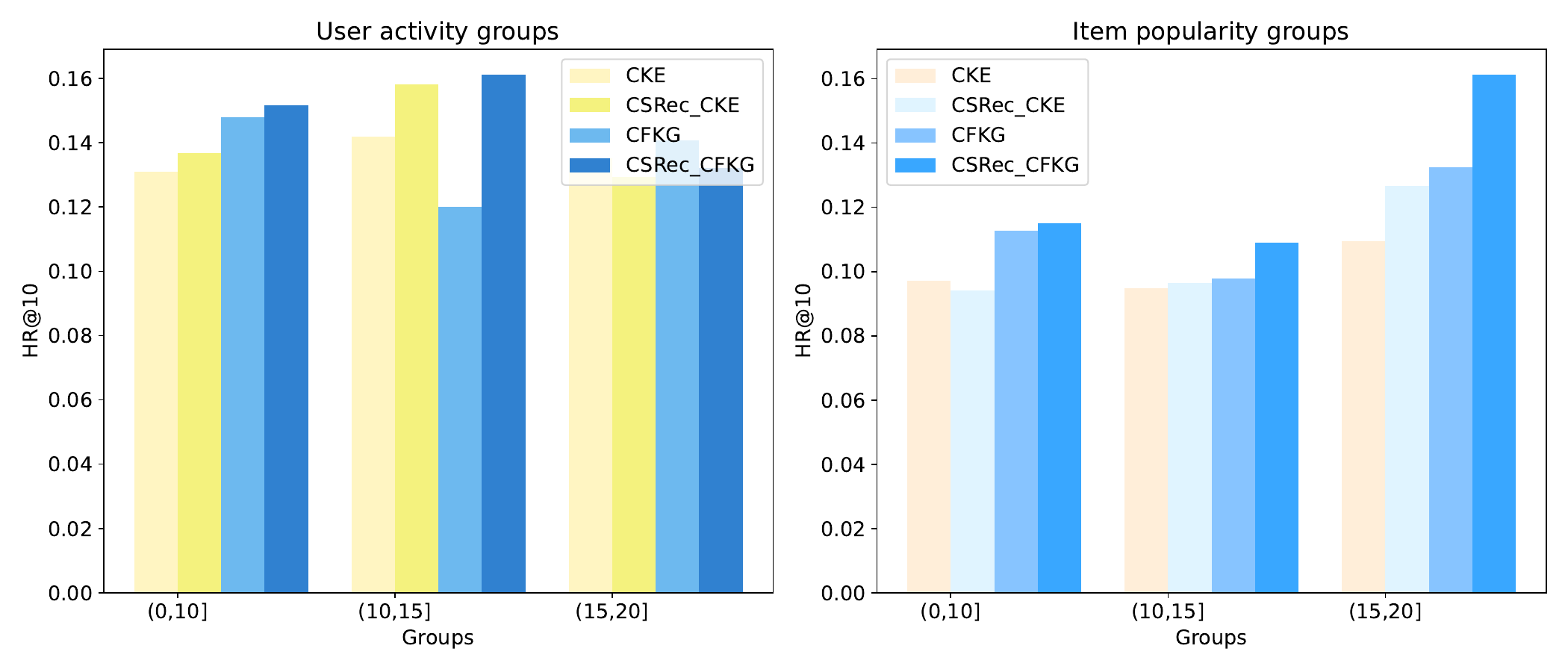}
\caption{Cold start Experiment results.} \label{fig3}
\end{figure}
To evaluate the performance of our proposed CSRec in cold-start scenarios, we compared the models based on CSRec and base models from the perspectives of cold users and cold items. Specifically, we grouped users and items based on their interaction counts and computed evaluation metrics within different groups. The experimental results under cold-start settings are presented in Figure~\ref{fig3}. From the results, several observations can be made. Firstly, in most cold user groups, the CSRec-based methods consistently exhibit performance improvements over the base models. Particularly noteworthy is the case of CSRec(CFKG), which achieves a relative improvement of 34\% in HR@10 compared to CFKG in the [10,15] group. Secondly, in the cold item experiments, the CSRec-based models consistently outperform the base models in almost all groups. Notably, CSRec(CFKG) achieves a performance improvement of approximately 21\% in the (15,20] group compared to the base model. Additionally, we observed a performance decrease in the extremely cold item group for the CSRec(CKE) models. We speculate that this could be attributed to the existence of popularity bias in common sense, making it challenging for these less frequent items to be captured by common sense.
\subsection{Case Study}
\begin{figure}[t]
\includegraphics[width=\textwidth]{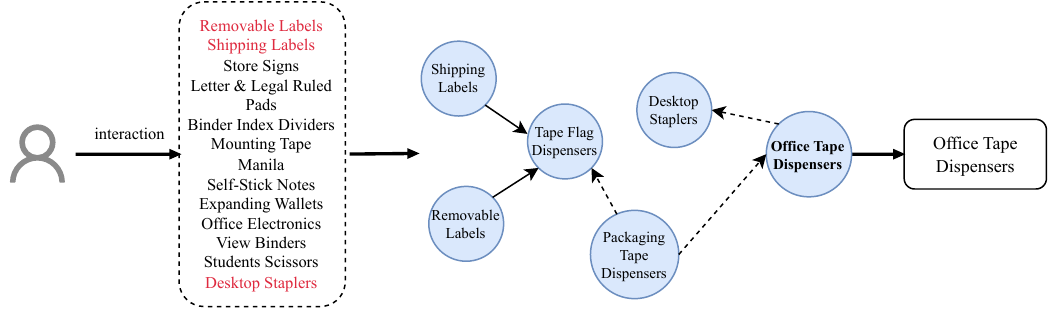}
\caption{Case study results.} \label{fig4}
\end{figure}

We present a case study to illustrate how CSRec leverages common sense in the recommendation process, as depicted in Figure~\ref{fig4}. On the left side are items previously interacted with by the user, where "Removable Labels" and "Shipping Labels" exhibit a complementary relationship. Additionally, "Tape Flag Dispensers" can be considered as an alternative to "Packaging Tape Dispensers," while the target item, "Office Tape Dispensers," is also related to "Packaging Tape Dispensers." Consequently, the model can effectively capture the target item with the aid of common sense. Experimental results demonstrate that the CSRec (CKE) model can recall the target item into the top 10, while other models fail to achieve this. This intuitively showcases the effectiveness of common sense-enhanced recommendations.
\section{Conclusion}
In this paper, we introduce CSRec, a pioneering knowledge-based recommendation framework that seamlessly integrates common sense, providing a valuable supplement to conventional metadata-based knowledge graphs. The inclusion of common sense, characterized by its generality and universality, addresses the limitations of metadata-based knowledge graphs, particularly their reliance on item attributes and interaction data. Leveraging the efficiency of large language models in acquiring world knowledge, our proposed CSRec framework offers a novel perspective for modeling user preferences.
To bridge the knowledge gap between common sense-based and metadata-based knowledge graphs, we introduce a knowledge fusion approach grounded in mutual information maximization theory. This method significantly enhances the effectiveness of existing knowledge-based recommendation models by offering a more comprehensive understanding of user preferences.
Experimental results on public datasets affirm the superior performance of CSRec compared to traditional knowledge-based recommendation models. The promising outcomes highlight the potential of incorporating common sense in recommendation systems, showcasing the adaptability and efficacy of our proposed framework. The successful integration of common sense in CSRec not only addresses the data sparsity issue but also demonstrates the versatility of this approach in enhancing recommendation model performance.
\bibliographystyle{splncs04}
\bibliography{reference}

\begin{thebibliography}{10}
\providecommand{\url}[1]{\texttt{#1}}
\providecommand{\urlprefix}{URL }
\providecommand{\doi}[1]{https://doi.org/#1}

\bibitem{ai2018learning}
Ai, Q., Azizi, V., Chen, X., Zhang, Y.: Learning heterogeneous knowledge base embeddings for explainable recommendation. Algorithms  \textbf{11}(9), ~137 (2018)

\bibitem{Bao_2023}
Bao, K., Zhang, J., Zhang, Y., Wang, W., Feng, F., He, X.: Tallrec: An effective and efficient tuning framework to align large language model with recommendation. In: Proceedings of the 17th ACM Conference on Recommender Systems. RecSys ’23, ACM (Sep 2023). \doi{10.1145/3604915.3608857}, \url{http://dx.doi.org/10.1145/3604915.3608857}

\bibitem{brown2020language}
Brown, T., Mann, B., Ryder, N., Subbiah, M., Kaplan, J.D., Dhariwal, P., Neelakantan, A., Shyam, P., Sastry, G., Askell, A., et~al.: Language models are few-shot learners. Advances in neural information processing systems  \textbf{33},  1877--1901 (2020)

\bibitem{covington2016deep}
Covington, P., Adams, J., Sargin, E.: Deep neural networks for youtube recommendations. In: Proceedings of the 10th ACM conference on recommender systems. pp. 191--198 (2016)

\bibitem{gao2023chatrec}
Gao, Y., Sheng, T., Xiang, Y., Xiong, Y., Wang, H., Zhang, J.: Chat-rec: Towards interactive and explainable llms-augmented recommender system (2023)

\bibitem{guo2020survey}
Guo, Q., Zhuang, F., Qin, C., Zhu, H., Xie, X., Xiong, H., He, Q.: A survey on knowledge graph-based recommender systems. IEEE Transactions on Knowledge and Data Engineering  \textbf{34}(8),  3549--3568 (2020)

\bibitem{he2017neural}
He, X., Liao, L., Zhang, H., Nie, L., Hu, X., Chua, T.S.: Neural collaborative filtering. In: Proceedings of the 26th international conference on world wide web. pp. 173--182 (2017)

\bibitem{hou2023deep}
Hou, X., Wang, Z., Liu, Q., Qu, T., Cheng, J., Lei, J.: Deep context interest network for click-through rate prediction (2023)

\bibitem{liu2023chatgpt}
Liu, J., Liu, C., Lv, R., Zhou, K., Zhang, Y.: Is chatgpt a good recommender? a preliminary study. arXiv preprint arXiv:2304.10149  (2023)

\bibitem{liu2023once}
Liu, Q., Chen, N., Sakai, T., Wu, X.M.: Once: Boosting content-based recommendation with both open- and closed-source large language models (2023)

\bibitem{logeswaran2018efficient}
Logeswaran, L., Lee, H.: An efficient framework for learning sentence representations. arXiv preprint arXiv:1803.02893  (2018)

\bibitem{openai2023gpt4}
OpenAI: Gpt-4 technical report (2023)

\bibitem{rendle2012bpr}
Rendle, S., Freudenthaler, C., Gantner, Z., Schmidt-Thieme, L.: Bpr: Bayesian personalized ranking from implicit feedback. arXiv preprint arXiv:1205.2618  (2012)

\bibitem{sarwar2001item}
Sarwar, B., Karypis, G., Konstan, J., Riedl, J.: Item-based collaborative filtering recommendation algorithms. In: Proceedings of the 10th international conference on World Wide Web. pp. 285--295 (2001)

\bibitem{sun2019research}
Sun, Z., Guo, Q., Yang, J., Fang, H., Guo, G., Zhang, J., Burke, R.: Research commentary on recommendations with side information: A survey and research directions. Electronic Commerce Research and Applications  \textbf{37},  100879 (2019)

\bibitem{touvron2023llama}
Touvron, H., Lavril, T., Izacard, G., Martinet, X., Lachaux, M.A., Lacroix, T., Rozi{\`e}re, B., Goyal, N., Hambro, E., Azhar, F., et~al.: Llama: Open and efficient foundation language models. arXiv preprint arXiv:2302.13971  (2023)

\bibitem{wang2019knowledge}
Wang, H., Zhao, M., Xie, X., Li, W., Guo, M.: Knowledge graph convolutional networks for recommender systems. In: The world wide web conference. pp. 3307--3313 (2019)

\bibitem{wang2019kgat}
Wang, X., He, X., Cao, Y., Liu, M., Chua, T.S.: Kgat: Knowledge graph attention network for recommendation. In: Proceedings of the 25th ACM SIGKDD international conference on knowledge discovery \& data mining. pp. 950--958 (2019)

\bibitem{yu2013collaborative}
Yu, X., Ren, X., Gu, Q., Sun, Y., Han, J.: Collaborative filtering with entity similarity regularization in heterogeneous information networks. IJCAI HINA  \textbf{27} (2013)

\bibitem{zhang2016collaborative}
Zhang, F., Yuan, N.J., Lian, D., Xie, X., Ma, W.Y.: Collaborative knowledge base embedding for recommender systems. In: Proceedings of the 22nd ACM SIGKDD international conference on knowledge discovery and data mining. pp. 353--362 (2016)

\bibitem{zhang2023recommendation}
Zhang, J., Xie, R., Hou, Y., Zhao, W.X., Lin, L., Wen, J.R.: Recommendation as instruction following: A large language model empowered recommendation approach (2023)

\bibitem{zhang2022opt}
Zhang, S., Roller, S., Goyal, N., Artetxe, M., Chen, M., Chen, S., Dewan, C., Diab, M., Li, X., Lin, X.V., et~al.: Opt: Open pre-trained transformer language models. arXiv preprint arXiv:2205.01068  (2022)

\end{thebibliography}
%




\end{document}